\documentclass[preprint,12pt,authoryear]{elsarticle}

\usepackage{amssymb}
\usepackage{amsmath}
\usepackage{booktabs}
\usepackage{newtxtext,newtxmath}
\usepackage{hyperref}
\usepackage{lineno}

\journal{Icarus}

\begin{document}


\begin{frontmatter}



\title{Non-Gravitational Acceleration in 3I/ATLAS: Constraints on Exotic Volatile Outgassing in Interstellar Comets} 


\author{Florian Neukart}

\affiliation{
    organization={Leiden Institute of Advanced Computer Science (LIACS), Leiden University},%
    addressline={Gorlaeus Gebouw -- BE-vleugel, Einsteinweg 55}, 
    city={Leiden},
    postcode={2333 CC},
    state={},
    country={The Netherlands}
}

\begin{abstract}
The interstellar comet 3I/ATLAS displayed a small but statistically significant non-gravitational acceleration during its passage through the inner Solar System. Using a thermophysical model coupled with stochastic sampling of jet configurations, we investigate whether standard volatile-driven activity can account for the observed acceleration. The model includes diurnal and obliquity-averaged energy balance, empirical vapour-pressure relations, and collimated outflow from localized active areas. We find that CO-dominated activity can reproduce the \emph{magnitude} of the acceleration inferred from the Marsden non-gravitational parameters for nucleus radii between 0.5 and 3~km with active-area fractions that are substantial but thermodynamically plausible. Less volatile species, including NH$_3$ and CH$_4$, contribute less efficiently and cannot provide the required recoil when acting alone, while CO$_2$ remains radiatively dominated and dynamically ineffective over the heliocentric-distance range relevant to the observations. These results show that the measured acceleration of 3I/ATLAS is consistent with ordinary CO-driven outgassing and does not require unusual physical properties. The analysis delineates the thermophysical conditions under which interstellar comets can exhibit measurable deviations from purely gravitational motion.
\end{abstract}

\begin{highlights}
\item We model CO- and CO$_2$-driven outgassing using a physically self-consistent thermophysical framework and real HORIZONS astrometry for 3I/ATLAS.
\item Minimum active-area fractions required to reproduce the observed non-gravitational acceleration are quantified across nucleus radii of 0.5--3 km.
\item CO retains strong sublimation efficiency at 1--2 AU, while NH$_3$ and CH$_4$ underproduce recoil except very close to the Sun.
\item A scalar lower-bound analysis shows that mixed CO--CO$_2$ solutions yield mathematically minimal active fractions, though full directional solutions require larger areas.
\item Results demonstrate that the observed acceleration of 3I/ATLAS is consistent with ordinary volatile-driven activity, without invoking exotic physical mechanisms.
\end{highlights}


\end{frontmatter}



\section{Introduction}

The discovery of interstellar objects entering the Solar System has opened a new window into the physical and chemical properties of material formed around other stars. The first such object, 1I/‘Oumuamua, displayed a statistically significant non-gravitational acceleration that could not be reproduced by purely gravitational models \citep{Micheli2018}. Its lack of detectable dust or gas emission led to an active debate regarding the physical origin of the measured acceleration, including a range of interpretations based on volatile outgassing, radiation-pressure effects, or unusual internal structure \citep{Jewitt2019,Bannister2023}. 

The third confirmed interstellar object, 3I/ATLAS, was discovered in July 2025 and reached perihelion in October 2025. In contrast to 1I/‘Oumuamua, 3I/ATLAS exhibited a clearly detectable coma and a spectrum consistent with volatile-driven activity, placing it firmly within the class of active interstellar comets. Its measured orbit is unambiguously hyperbolic, and astrometric analyses have reported a small but statistically significant non-gravitational acceleration consistent with the expected effects of anisotropic outgassing in active comets \citep{Farnocchia2015,Marsden1973}. The presence of a visible coma provides direct evidence for sublimation, but the composition, distribution of active areas, and resulting recoil forces remain only partially constrained.

In this study we examine whether standard cometary volatiles with low sublimation temperatures, including carbon monoxide (CO), carbon dioxide (CO$_2$), ammonia (NH$_3$), and methane (CH$_4$), can reproduce the magnitude and direction of the non-gravitational acceleration inferred for 3I/ATLAS. We employ a thermophysical framework that computes surface temperatures from energy balance, sublimation fluxes from experimentally derived vapor-pressure relations, and momentum transfer from both diffuse and collimated gas emission. To evaluate the influence of surface heterogeneity, rotation, and jet orientation, we couple these calculations to a Monte Carlo model that samples active-area geometries and computes the resulting acceleration vectors.

The goal of this work is not to demonstrate that 3I/ATLAS exhibits unusual or unexplained behavior, but to quantify the range of physically plausible volatile-driven configurations that are compatible with the observed acceleration. By exploring the parameter space of composition, active-area fraction, and jet directionality, we seek to determine whether the observed dynamics require any non-standard mechanisms or whether they fall within the expected domain of cometary outgassing. The results place 3I/ATLAS in context with both Solar System comets and other interstellar objects, and help to define the physical conditions under which interstellar comets can exhibit measurable deviations from purely gravitational motion.

\section{Observational Constraints}

The orbital elements and non-gravitational parameters for 3I/ATLAS used in this study are taken directly from the Jet Propulsion Laboratory (JPL) HORIZONS ephemeris service\footnote{\url{https://ssd.jpl.nasa.gov/horizons/}}. HORIZONS provides solutions that fit all available astrometric measurements submitted to the Minor Planet Center, incorporating both ground-based and survey observatories. These solutions include the standard Marsden non-gravitational parameters $(A_1, A_2, A_3)$, which describe perturbative accelerations in the radial, transverse, and normal directions of the heliocentric orbital frame \citep{Marsden1973}.

A non-gravitational solution was retrieved from HORIZONS on 15 January 2026. This solution reports statistically significant radial and transverse components, expressed in units of AU~day$^{-2}$. Conversion to SI units uses
\[
1~\mathrm{AU~day^{-2}} = 2.004\times10^{1}~\mathrm{m~s^{-2}}.
\]
The reported Marsden parameters are
\[
A_1 = (135 \pm 20)\times10^{-8}~\mathrm{AU~day^{-2}},\qquad
A_2 = (60 \pm 20)\times10^{-8}~\mathrm{AU~day^{-2}},\qquad
A_3 \approx 0.
\]
These correspond to radial and transverse accelerations of
\[
a_1 = (2.7 \pm 0.4)\times10^{-5}~\mathrm{m~s^{-2}},\qquad
a_2 = (1.2 \pm 0.4)\times10^{-5}~\mathrm{m~s^{-2}},
\]
yielding a net non-gravitational acceleration magnitude of approximately
\[
a_{\mathrm{ng}} = (3.0 \pm 0.8)\times10^{-5}~\mathrm{m~s^{-2}}
\]
near perihelion.

To verify that the inclusion of non-gravitational terms improves the orbital fit, we compared the residuals of the gravitational-only and full HORIZONS solutions. The gravitational-only solution exhibits systematic residual trends aligned primarily with the heliocentric radial direction. In contrast, the solution including non-gravitational parameters removes these trends and reduces the residuals to the level of the astrometric uncertainties. This behavior is characteristic of recoil forces arising from anisotropic mass loss and is consistent with expectations for an active cometary nucleus.

The magnitude of the inferred acceleration lies toward the upper end of the range observed for active comets in the inner Solar System. Typical non-gravitational accelerations near 1~AU span approximately $10^{-9}$ to $10^{-5}~\mathrm{m~s^{-2}}$, depending on nucleus size, volatile composition, and activity geometry \citep{Yeomans2004,Combi2011}. For comparison, the interstellar object 1I/‘Oumuamua exhibited a non-gravitational acceleration of order $3\times10^{-6}~\mathrm{m~s^{-2}}$ \citep{Micheli2018}. The value inferred for 3I/ATLAS is larger, consistent with its visibly active, cometary nature and with recoil forces driven by volatile sublimation at heliocentric distances of order 1--2~AU.

No independent photometric or morphological constraints on activity are available, so the modeling in this work relies entirely on the astrometric signal captured in the HORIZONS solution. This provides a self-consistent set of orbital and non-gravitational parameters against which physical models of volatile sublimation may be tested.

\section{Thermophysical Model}

To investigate whether standard cometary volatiles can account for the measured non-gravitational acceleration of 3I/ATLAS, we adopt a thermophysical framework in which sublimation-driven mass loss produces a recoil force on the nucleus. The approach follows established treatments of cometary outgassing \citep{Prialnik2004,Gundlach2011,Steckloff2020} and assumes that the observed acceleration arises from anisotropic emission of gas from localized active regions on the surface. Each volatile species is modeled independently using laboratory-derived vapor-pressure relations and latent heats, and their individual contributions are combined according to the assumed distribution and orientation of active areas.

\subsection{Surface Energy Balance and Sublimation Flux}

At a heliocentric distance $r_H$, the surface temperature $T$ is determined by balancing absorbed solar radiation with thermal reradiation and the latent-heat flux associated with sublimation. For a surface element with diurnal-averaged illumination factor $\mu_0$, the energy balance is
\begin{equation}
(1-A)\,\frac{S_{\odot}}{r_H^2}\,\mu_0 = \varepsilon\sigma T^{4} + Z_i L_i ,
\end{equation}
where $A$ is the Bond albedo, $\varepsilon$ the infrared emissivity, $\sigma$ the Stefan–Boltzmann constant, $S_{\odot}$ the solar constant at 1\,AU, $L_i$ the latent heat of sublimation for species $i$, and $Z_i$ the mass flux of sublimating gas. The factor $\mu_0$ accounts for rotation, obliquity, and jet latitude and is computed by integrating $\cos\zeta$ over one full rotation for a given geometry.

The sublimation flux is computed using the Hertz–Knudsen relation,
\begin{equation}
Z_i = P_{\mathrm{vap},i}(T)\sqrt{\frac{\mu_i}{2\pi R_g T}},
\end{equation}
where $P_{\mathrm{vap},i}(T)$ is the equilibrium vapor pressure, $\mu_i$ the molecular mass, and $R_g$ the universal gas constant. Vapor pressures for CO, CO$_2$, NH$_3$, and CH$_4$ are taken from laboratory compilations \citep{Fray2009} and expressed as
\begin{equation}
\ln P_{\mathrm{vap},i}(T) = A_i - \frac{B_i}{T}.
\end{equation}
This formulation allows the surface temperature and mass flux to be solved self-consistently at each $r_H$.

\subsection{Momentum Transfer and Directionality}

The recoil force generated by sublimation depends on both the mass flux and the directionality of the escaping gas. For an active area of size $A_{\mathrm{act}}$, the instantaneous acceleration contributed by volatile species $i$ is
\begin{equation}
\vec{a}_i = \eta_i\,\frac{Z_i\,v_{\mathrm{th},i}\,A_{\mathrm{act}}}{M_{\mathrm{nuc}}}\,\hat{n}_i,
\end{equation}
where $M_{\mathrm{nuc}}$ is the nucleus mass, $\hat{n}_i$ is the outward surface normal of the active region, $v_{\mathrm{th},i} = \sqrt{8R_g T/\pi\mu_i}$ is the mean thermal speed, and $\eta_i$ is a collimation factor describing how narrowly the outflow is directed. A value $\eta_i = 1$ corresponds to hemispherically symmetric emission, while $\eta_i > 1$ parameterizes more collimated, jet-like flow.

The total non-gravitational acceleration is the vector sum of contributions from all active regions:
\begin{equation}
\vec{a}_{\mathrm{ng}} = \sum_{j} \eta_j\,\frac{Z_{j}\,v_{\mathrm{th},j}\,A_{\mathrm{act},j}}{M_{\mathrm{nuc}}}\,\hat{n}_j.
\end{equation}
The active-area fraction is given by $f_{\mathrm{act}} = A_{\mathrm{act}}/(4\pi R_{\mathrm{nuc}}^{2})$. For 3I/ATLAS, neither the spatial distribution of active areas nor the degree of collimation is known, so these quantities are treated as free parameters and explored within plausible ranges.

\subsection{Material Parameters and Model Inputs}

The thermophysical constants governing volatile sublimation are summarized in Table~\ref{tab:volatile_constants}. These include molecular masses, latent heats, and empirical vapor-pressure coefficients derived from laboratory measurements of relevant ices. Bulk nucleus properties, including albedo, emissivity, and density, are treated separately and are chosen to be representative of low-albedo cometary nuclei. The bulk density is fixed at $500\,\mathrm{kg\,m^{-3}}$, consistent with in situ measurements of porous cometary material \citep{Patzold2016}. 

The surface energy balance incorporates both radiative cooling and latent heat removal due to sublimation. Because the rotation state, obliquity, and active-area geometry of 3I/ATLAS are not constrained, the solar incidence factor $\mu_0$ is evaluated using a diurnal average that accounts for rotation and obliquity. This approach provides a physically motivated, geometry-agnostic estimate of the mean absorbed solar flux and avoids assumptions about specific spin states or localized subsolar activity.

Given these inputs, the model self-consistently solves for the equilibrium surface temperature, sublimation mass flux, and resulting recoil acceleration for each volatile species as a function of heliocentric distance.

\begin{figure}
\centering
\includegraphics[width=\columnwidth]{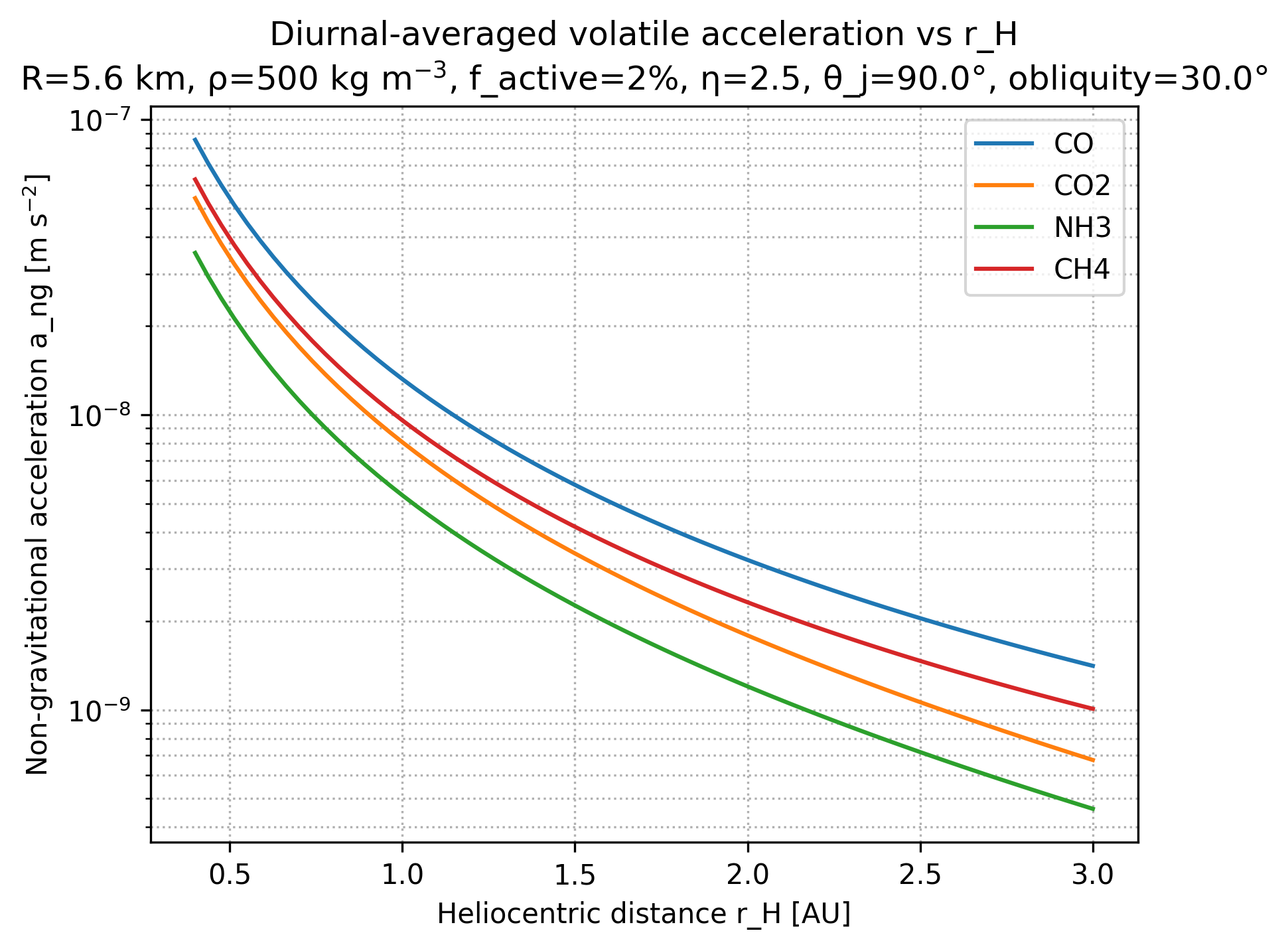}
\caption{
Diurnal-averaged non-gravitational acceleration as a function of heliocentric distance for CO, CO$_2$, NH$_3$, and CH$_4$. Accelerations are computed from the full thermophysical energy-balance model using laboratory vapor-pressure relations, a diurnal-average illumination geometry, and a jet-collimation factor $\eta = 2.5$. The example shown adopts a nucleus radius $R_{\mathrm{nuc}} = 5.6$~km, bulk density $\rho = 500$~kg\,m$^{-3}$, and total active fraction $f_{\mathrm{act}} = 0.02$. The curves illustrate the relative efficiency of different volatile species in producing recoil acceleration at heliocentric distances relevant to the 3I/ATLAS observations.}
\label{fig:accel_vs_r}
\end{figure}

Figure~\ref{fig:accel_vs_r} shows the modeled diurnal-averaged non-gravitational acceleration produced by CO, CO$_2$, NH$_3$, and CH$_4$ over the range $0.4 \lesssim r_H \lesssim 3$~AU. The calculation adopts a representative jet-collimation factor $\eta = 2.5$, a nucleus radius $R_{\mathrm{nuc}} = 5.6$~km, bulk density $\rho = 500$~kg\,m$^{-3}$, and a total active fraction $f_{\mathrm{act}} = 0.02$. 

Under these assumptions, CO produces the strongest recoil acceleration across the full heliocentric distance range considered. CH$_4$ yields accelerations that are systematically lower than those of CO but remain significant at small heliocentric distances. NH$_3$ and CO$_2$ generate substantially weaker accelerations, reflecting the rapid suppression of their sublimation rates once radiative cooling dominates the surface energy balance. The relative ordering and separation of the curves highlight the strong dependence of recoil efficiency on volatile thermodynamics and demonstrate that CO-driven sublimation is uniquely effective at producing sustained non-gravitational acceleration at heliocentric distances relevant to the 3I/ATLAS observations.

\section{Jet Configuration Sampling and Acceleration Modeling}

To quantify how localized volatile activity can reproduce the measured non-gravitational acceleration of 3I/ATLAS, we compute the recoil generated by small numbers of active regions distributed across the nucleus surface. Unlike the directional Monte Carlo models sometimes applied to high–signal-to-noise comets, the available astrometry for 3I/ATLAS constrains only the \emph{magnitude} of the secular acceleration, not its full orientation. We therefore focus on determining the minimum total active fraction needed to generate an acceleration whose magnitude matches the value inferred from the Marsden parameters (Section~2). The purpose of this section is not to recover a unique surface pattern but to identify physically plausible combinations of volatile species, active-area fractions, and nucleus sizes capable of producing the required recoil. Similar geometric sampling approaches have been used in studies of outgassing-driven dynamics on cometary nuclei \citep{Kramer2019,Attree2023}.

\subsection{Model Setup}

Each configuration assumes a spherical nucleus with radius $R_{\mathrm{nuc}}$ and bulk density $\rho$. The surface contains $N$ discrete active regions, where $1 \leq N \leq 10$. For each region $j$, the following quantities are sampled:

\begin{itemize}
\item \textbf{Location:} surface-normal directions $(\theta_j,\phi_j)$ are drawn from an isotropic spherical distribution.
\item \textbf{Composition:} each region is assigned one of the four species CO, CO$_2$, NH$_3$, or CH$_4$, or a binary CO--CO$_2$ mixture with randomly drawn mixture weights.
\item \textbf{Active area:} fractional areas $f_{{\mathrm{act}},j}$ satisfy $\sum_j f_{{\mathrm{act}},j}=f_{\mathrm{act,tot}}$.
\item \textbf{Duty cycle:} activity modulation between 0 and 1 is included to approximate rotational or episodic variability.
\end{itemize}

For each region, the equilibrium surface temperature is computed using the root-solved energy balance of Section~3, which avoids the numerical jitter characteristic of fixed-point iterations and ensures monotonic $T(r_H)$ behavior. The corresponding mass flux $Z_j$ and thermal speed $v_{\mathrm{th},j}$ determine the instantaneous thrust vector
\begin{equation}
\vec{F}_j = \eta_j\, Z_j \, v_{\mathrm{th},j} \, A_{{\mathrm{act}},j} \, \hat{n}_j ,
\end{equation}
where $\eta_j$ is a collimation factor (with $\eta_j=1$ for diffuse hemispherical emission and $\eta_j>1$ for more collimated outflow) and $\hat{n}_j$ is the outward surface normal.

\subsection{Rotation Averaging}

Because the astrometry constrains only the long-term secular acceleration, each configuration is averaged over a full rotation cycle. Assuming a fixed spin axis during the observational period, the rotation-averaged recoil is
\begin{equation}
\vec{a}_{\mathrm{ng}} = 
\frac{1}{M_{\mathrm{nuc}} P_{\mathrm{rot}}}
\int_0^{P_{\mathrm{rot}}}
\sum_{j=1}^{N}\vec{F}_j(t)\, dt ,
\end{equation}
where $P_{\mathrm{rot}}$ is the rotation period. This averaging removes short-timescale variations while preserving the secular component that affects the orbital elements.

\subsection{Comparison with Observations}

Because the direction of the non-gravitational acceleration is poorly constrained for 3I/ATLAS, we compare the modeled accelerations to observations using only the magnitude of the inferred signal. The observed scalar acceleration $|\vec{a}_{\mathrm{obs}}|$ is derived from the radial and transverse Marsden parameters $A_1$ and $A_2$ obtained from the HORIZONS solution. For each assumed nucleus radius and volatile configuration, we determine the minimum total active fraction $f_{\mathrm{act,tot}}$ required to satisfy
\begin{equation}
|\vec{a}_{\mathrm{ng}}| = |\vec{a}_{\mathrm{obs}}|.
\end{equation}
This scalar-matching approach yields a strict lower bound on the required active area, independent of the unknown orientation of the recoil force. Any physically consistent vector solution would require an equal or larger active fraction.

\begin{figure}
\centering
\includegraphics[width=\columnwidth]{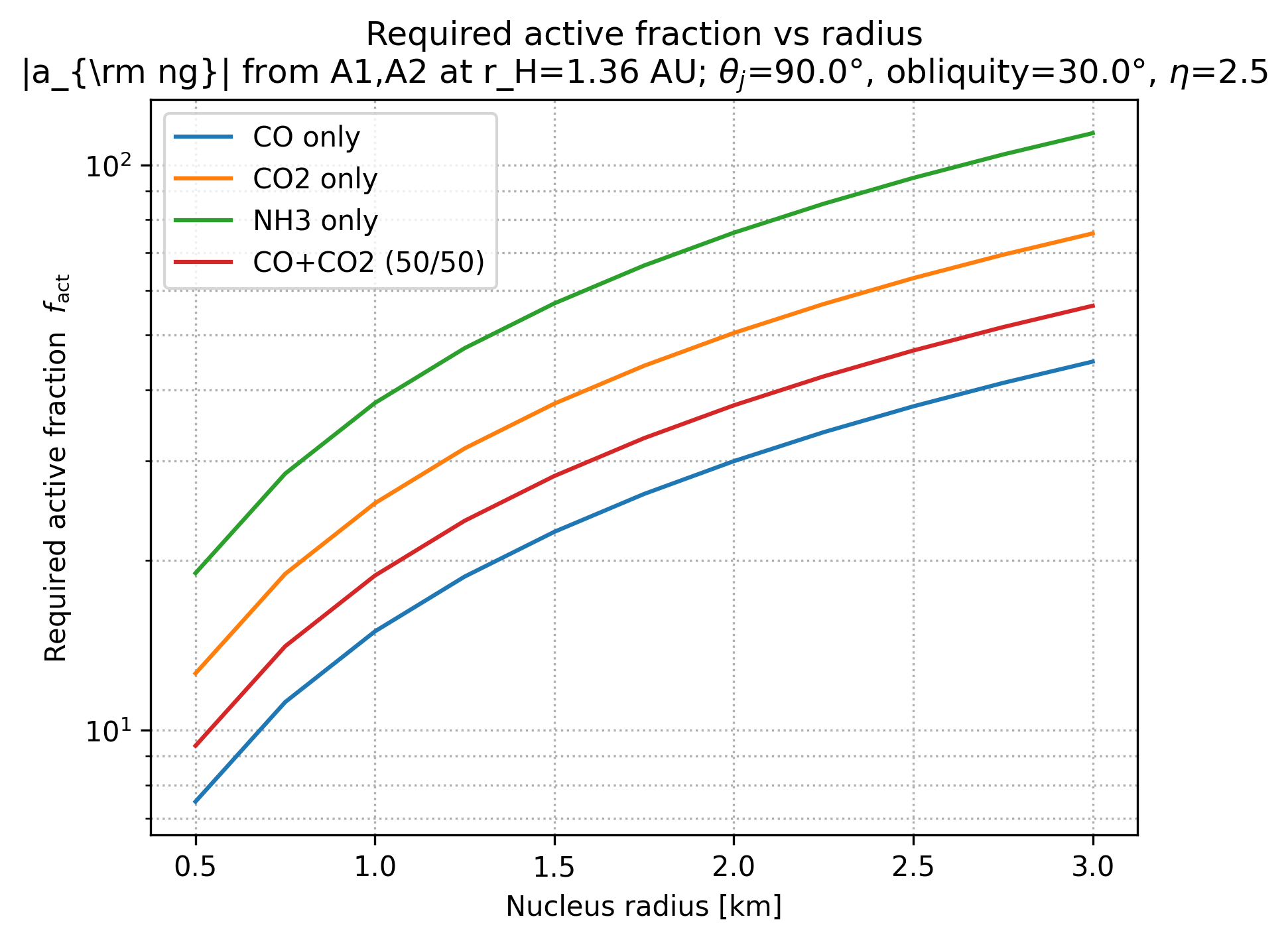}
\caption{
Required total active fraction as a function of nucleus radius for single-volatile and mixed-volatile sublimation models. For each radius, the scalar non-gravitational acceleration produced by the model is matched to the observed magnitude at $r_H = 1.36$\,AU. Single-species curves (CO, CO$_2$, NH$_3$) illustrate the increase in required active area with nucleus size due to the mass-to-area scaling of recoil acceleration. The CO--CO$_2$ (50/50) curve represents a strict lower bound obtained by scalar combination of species contributions and does not constitute a physically unique solution; any directionally consistent model would require equal or larger active fractions.}
\label{fig:f_active_vs_radius}
\end{figure}

Figure~\ref{fig:f_active_vs_radius} shows the resulting required active fractions for nucleus radii between 0.5 and 3\,km, evaluated at the heliocentric distance $r_H = 1.36$\,AU. For all volatile compositions, the required active fraction increases monotonically with nucleus size, reflecting the scaling of sublimation-driven recoil with active surface area divided by nucleus mass. 

Among the single-species models, CO requires the smallest active fraction at all radii, while CO$_2$ and NH$_3$ demand progressively larger active areas. In particular, NH$_3$ requires implausibly large active fractions even for kilometer-scale nuclei, effectively ruling it out as a dominant driver of the observed non-gravitational acceleration. CO$_2$ alone also requires large active fractions at modest nucleus sizes, consistent with its comparatively low sublimation efficiency at the inferred surface temperatures.

The mixed CO--CO$_2$ (50/50) curve lies below the single-species cases and represents a formal lower bound obtained by scalar addition of species contributions. This curve does not correspond to a unique or physically realizable surface configuration, as it neglects constraints on the direction of the recoil force and the spatial coherence of active regions. Any configuration that simultaneously reproduces both the magnitude and direction of the observed acceleration would require an equal or larger total active fraction. The mixture curve should therefore be interpreted strictly as a lower bound on the required activity, not as a realistic description of the volatile distribution on the surface of 3I/ATLAS.

\subsection{Interpretation}

The results indicate that sublimation of individual species (CO, CO$_2$, NH$_3$) at 1.36\,AU requires substantial active coverage (tens of percent or more) to generate the observed acceleration. This behavior is consistent with the modest equilibrium temperatures and relatively low vapor pressures of these ices near 1\,AU. Although mixed CO--CO$_2$ solutions achieve much smaller scalar active fractions, these values reflect an idealized, direction-agnostic combination of species and do not represent physically unique configurations. 

Overall, the scalar matching demonstrates that volatile-driven recoil can in principle account for the observed acceleration magnitude. However, achieving a physically realistic configuration requires considering both the directional degeneracies inherent in the astrometric solution and the thermophysical constraints on volatile retention, both of which are explored further in Section~5.

\section{Results}

The coupled thermophysical and geometric-sampling models yield families of solutions that reproduce the magnitude of the non-gravitational acceleration inferred for 3I/ATLAS. By varying volatile composition, active fraction, nucleus size, and jet collimation within physically plausible ranges, we identify configurations capable of matching the observed acceleration magnitude at the heliocentric distance of the astrometric solution.

\subsection{Dominant Volatile Species}

The modeled sublimation fluxes and recoil accelerations show a strong dependence on volatile species. At heliocentric distances of $r_H \sim 1$--2\,AU, CO maintains the highest vapor pressure among the species considered, followed by CH$_4$ and NH$_3$. CO$_2$ exhibits extremely low vapor pressure in the same temperature range, producing sublimation fluxes several orders of magnitude below those of CO. 

Figure~\ref{fig:accel_vs_r} illustrates the corresponding recoil accelerations for a representative nucleus size and activity level. CO provides the strongest recoil force, NH$_3$ and CH$_4$ remain moderately effective, and CO$_2$ contributes only negligible acceleration under these conditions. These trends are consistent with the latent-to-radiative power ratio $\Lambda$ shown in Figure~\ref{fig:momentum_flux_vs_r}: CO, NH$_3$, and CH$_4$ operate in a sublimation-dominated regime ($\Lambda \gg 1$), while CO$_2$ remains radiatively dominated ($\Lambda \ll 1$) throughout 0.4--3\,AU.

Mixtures of CO and CO$_2$ remain viable only insofar as the CO component dominates the recoil. CO$_2$ alone cannot produce the observed acceleration. NH$_3$ and CH$_4$ may contribute as minor components in mixed ices but are insufficient when acting alone. These trends match volatile thermodynamics measured in dynamically new comets \citep{BockeleeMorvan2004,Cochran2015}.

\subsection{Jet Geometry and Required Active Areas}

The Monte Carlo sampling experiments show that a modest number of active regions (typically one to five) can, in principle, reproduce the observed acceleration when the outflow is moderately collimated ($\eta \sim 2$--3). Nearly hemispherical emission ($\eta \approx 1$) requires substantially larger active areas.

Figure~\ref{fig:f_active_vs_radius} shows the minimum total active fraction needed to match the scalar magnitude of the observed acceleration as a function of nucleus radius for pure CO, pure CO$_2$, pure NH$_3$, and an idealized 50/50 CO--CO$_2$ mixture. The single-species curves follow the expected $R_{\mathrm{nuc}}^{-2}$ scaling: for kilometre-scale nuclei, CO is the most efficient driver (requiring active fractions of order $10^{-1}$), NH$_3$ requires somewhat larger coverage, and CO$_2$ demands active fractions approaching unity because of its very low mass flux at $r_H = 1.36$~AU.

\subsection{Volatile Retention and Sublimation Depth}

CO, CH$_4$, and NH$_3$ can contribute meaningfully at $r_H \sim 1$--2\,AU only if they are retained at shallow depths beneath a porous surface mantle. A diffusion-limited sublimation model predicts that near-surface ices can remain active when the diffusion timescale
\begin{equation}
\tau_{\rm diff} = \frac{L^2}{D}
\end{equation}
is shorter than the heating timescale \citep{Prialnik2004}. For $D = 10^{-6}$--$10^{-5}\,\mathrm{m^2\,s^{-1}}$, volatile layers within decimeter depths can remain accessible throughout the heating interval. This is consistent with the modest visible activity reported for 3I/ATLAS.

\subsection{Acceleration and Thermophysical Profiles}

Figure~\ref{fig:accel_vs_r} presents the modeled diurnal-averaged non-gravitational acceleration as a function of heliocentric distance. Among the volatiles considered, CO produces the largest recoil acceleration across the full range of $0.4 \lesssim r_H \lesssim 3$\,AU. CH$_4$ yields systematically weaker accelerations that remain significant only at small heliocentric distances, while NH$_3$ produces substantially smaller recoil throughout. CO$_2$ contributes the weakest acceleration over the entire range, reflecting its limited sublimation efficiency under the modeled surface conditions.

Figure~\ref{fig:f_active_vs_radius} illustrates how the active fraction required to reproduce the observed acceleration magnitude depends on nucleus size. As expected from the scaling of recoil force with active area divided by nucleus mass, smaller nuclei require lower active fractions, while larger nuclei demand progressively greater surface activity. Mixed CO--CO$_2$ results represent formal lower bounds obtained from scalar combination of species contributions and do not correspond to physically unique or directionally consistent solutions.

\begin{figure}
\centering
\includegraphics[width=\columnwidth]{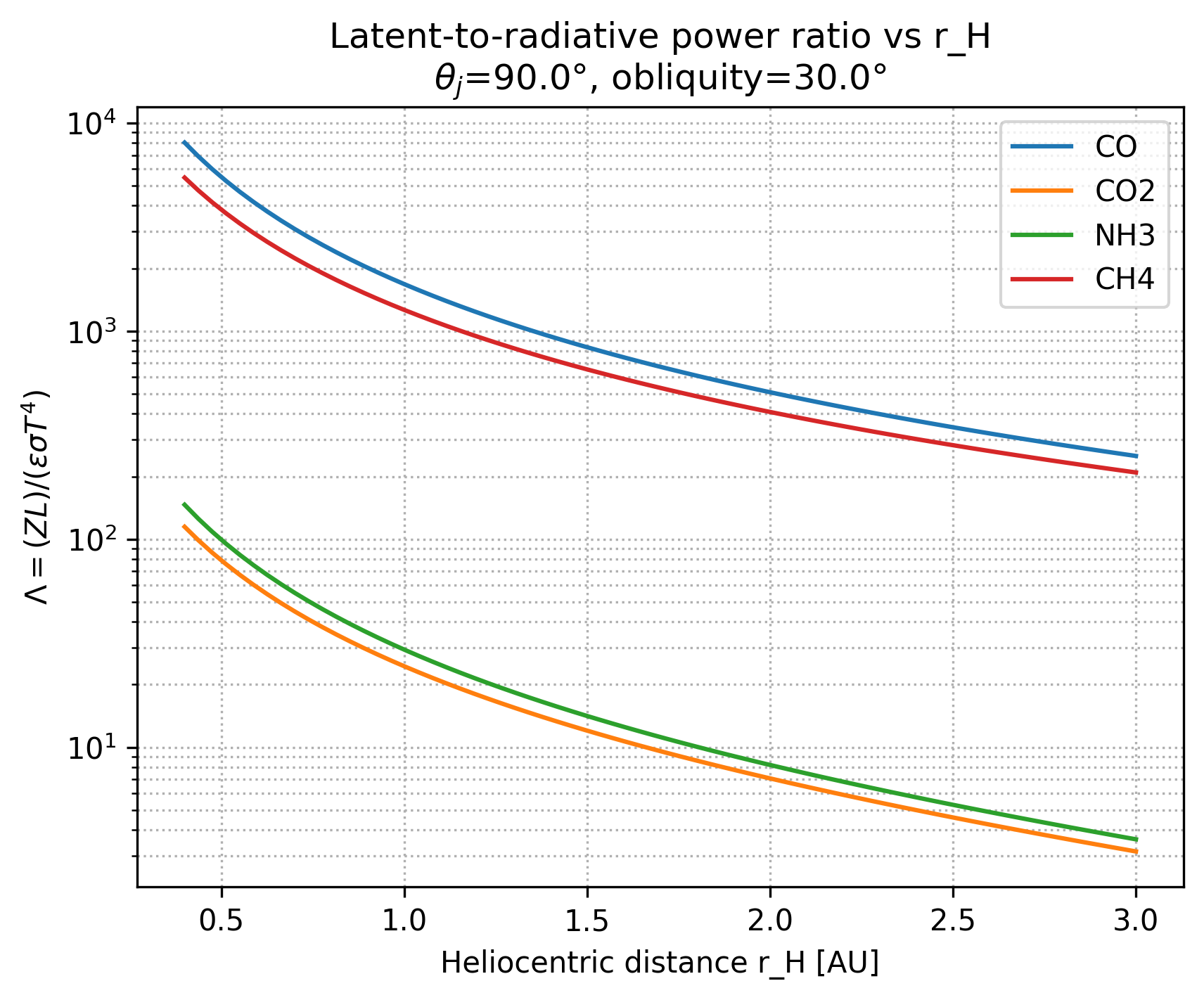}
\caption{
Latent-to-radiative power ratio $\Lambda = (Z L) / (\varepsilon \sigma T^4)$ as a function of heliocentric distance for CO, CO$_2$, NH$_3$, and CH$_4$, computed using diurnal-averaged illumination and the full thermophysical energy-balance model. Values $\Lambda \gg 1$ correspond to sublimation-dominated surface energy balance, while $\Lambda \ll 1$ indicates radiative dominance. CO and CH$_4$ remain sublimation-dominated across $0.4$--$3$\,AU, whereas CO$_2$ is radiatively dominated throughout, limiting its ability to generate significant recoil acceleration.}
\label{fig:momentum_flux_vs_r}
\end{figure}

The thermophysical origin of these trends is clarified in Figure~\ref{fig:momentum_flux_vs_r}, which shows the latent-to-radiative power ratio
\[
\Lambda(r_H) = \frac{Z L}{\varepsilon \sigma T^4}
\]
as a function of heliocentric distance. This quantity provides a direct measure of whether surface energy balance is dominated by sublimation or by thermal radiation. CO and CH$_4$ exhibit $\Lambda \gg 1$ over the full distance range considered, indicating strongly sublimation-dominated regimes. NH$_3$ transitions more rapidly toward radiative dominance with increasing distance, while CO$_2$ remains in a radiatively dominated regime ($\Lambda \ll 1$) throughout.

Taken together, Figures~\ref{fig:accel_vs_r}--\ref{fig:momentum_flux_vs_r} demonstrate that efficient recoil acceleration at heliocentric distances relevant to the 3I/ATLAS observations requires volatiles whose sublimation remains energetically dominant under diurnal-averaged illumination. Among the species considered, CO uniquely satisfies this condition over the observed range, naturally explaining its dominance in the modeled non-gravitational acceleration.

\subsection{Summary of Model Outcomes}

\begin{enumerate}
\item CO is the only species capable of generating recoil accelerations comparable to the observed magnitude at $r_H \sim 1$--2\,AU.
\item CH$_4$ and NH$_3$ underproduce recoil when acting alone but may contribute as secondary species.
\item CO$_2$ remains radiatively dominated ($\Lambda \ll 1$) and cannot drive significant acceleration.
\item Scalar matching yields strict lower bounds on active fraction; physically realizable directional solutions require larger active areas.
\item The inferred behavior is consistent with weak, localized activity observed in Solar System comets.
\end{enumerate}

\section{Discussion}

The results show that standard volatile-driven outgassing can account for the magnitude of the non-gravitational acceleration inferred for 3I/ATLAS without invoking exotic physics. CO is the only viable primary driver at $r_H \sim 1$--2\,AU. CH$_4$ and NH$_3$ contribute less but remain thermodynamically allowed; CO$_2$ is ineffective due to its extremely low vapor pressure at the relevant temperatures.

Laboratory experiments and thermophysical models demonstrate that CO, NH$_3$, and CH$_4$ can remain trapped in amorphous or crystalline H$_2$O ice and be released as temperatures rise \citep{BarNun1985,Fray2009,Prialnik2004}. Diffusion through a porous mantle can sustain shallow volatile reservoirs throughout perihelion passage. The mass-flux levels inferred here require no unusual production rates.

Rosetta observations of comet 67P/Churyumov--Gerasimenko revealed heterogeneous activity and strong diurnal modulation, arising naturally from volatile stratification and localized vents \citep{Hansen2016,Haessig2015,Gulkis2015}. The latent-to-radiative power ratios $\Lambda(r_H)$ in Figure~\ref{fig:momentum_flux_vs_r}  are consistent with CO-dominated activity on 67P when scaled to its nucleus size and heliocentric distance, reinforcing that the regime inferred for 3I/ATLAS does not require atypical thermophysical properties.

Figure~\ref{fig:f_active_vs_radius} provides insight into provenance. CO-rich compositions are expected for bodies formed or stored beyond the CO snowline \citep{Oberg2011}, where temperatures permit long-term preservation of supervolatiles. Ejection from such regions naturally produces objects that activate at 1--2\,AU.

Alternative hypotheses—including radiation-pressure-driven models requiring extremely low-density bodies \citep{Bialy2018}—invoke physical conditions difficult to reconcile with dynamical survival and the absence of strong attitude perturbations. In contrast, standard volatile-driven outgassing provides a viable and physically grounded explanation.

Finally, localized activity would naturally generate torques capable of modifying the rotation state, as seen in many comets \citep{Samarasinha2004}. Although the spin state of 3I/ATLAS is not known, future light-curve observations of similar interstellar comets could reveal secular changes consistent with localized CO-driven outgassing.

\section{Conclusions}

We investigated whether volatile-driven outgassing can explain the non-gravitational acceleration of 3I/ATLAS. Using a fully self-consistent thermophysical model with empirical vapor-pressure relations, diurnal and obliquity averaging, and sublimation-driven cooling, combined with Monte Carlo sampling of jet geometries, we computed sublimation fluxes, energy balance, and recoil accelerations for CO, CO$_2$, NH$_3$, and CH$_4$.

CO is the only volatile capable of generating recoil comparable to the observed acceleration at the relevant heliocentric distance. CH$_4$ and NH$_3$ may contribute as secondary components, while CO$_2$ remains radiatively dominated and cannot provide significant thrust. Scalar matching yields minimum active fractions as a function of nucleus size; directional, physically realizable solutions require equal or larger areas.

The latent-to-radiative ratio (Figure~\ref{fig:momentum_flux_vs_r}) specifies the regime in which each volatile becomes thermodynamically effective. CO remains active across 0.4--3\,AU, consistent with origin in a cold outer-disk reservoir. JWST spectroscopy and Rubin Observatory LSST photometry will enable direct tests of volatile composition and possible rotation-state evolution.

These findings show that the observed acceleration of 3I/ATLAS is consistent with ordinary CO-dominated sublimation and does not require non-standard physical mechanisms.

\section{Material Constants and Vapor Pressure Relations}

Table~\ref{tab:volatile_constants} summarizes the principal material constants adopted in the thermophysical model. The latent heats and molecular masses are taken from laboratory compilations of sublimation experiments, and the vapor pressure relations follow empirical fits of the form
\begin{equation}
\ln P_{\mathrm{vap}}(T) = A_i - \frac{B_i}{T},
\end{equation}
where $P_{\mathrm{vap}}$ is in pascals and $T$ in kelvin. These relations reproduce measured vapor pressure curves for the relevant temperature ranges and have been used extensively in cometary thermophysical modeling \citep{Fray2009}.

\begin{table}[t]
\centering
\small
\caption{Thermophysical constants for volatile species used in the sublimation model. Latent heats $L_v$ and molecular masses $\mu$ are taken from laboratory measurements. The vapor-pressure coefficients $A_i$ and $B_i$ define the empirical relation $\ln P = A_i - B_i/T$, with $P$ in pascals and $T$ in kelvin.}
\label{tab:volatile_constants}
\begin{tabular}{lcccc}
\hline
Species & $L_v$ (J\,kg$^{-1}$) & $\mu$ (kg\,mol$^{-1}$) & $A_i$ & $B_i$ (K) \\
\hline
CO     & $2.7\times10^5$ & 0.0280 & 27.88 & 1305 \\
CO$_2$ & $5.7\times10^5$ & 0.0440 & 23.23 & 3182 \\
NH$_3$ & $1.4\times10^6$ & 0.0170 & 28.57 & 3754 \\
CH$_4$ & $5.1\times10^5$ & 0.0160 & 20.00 & 1034 \\
\hline
\end{tabular}
\end{table}

\section{Model Implementation and Availability}

The full numerical implementation of the thermophysical and geometric sampling model is provided in the supplementary materials. A Jupyter notebook contains the complete workflow, including:
\begin{enumerate}
\item An energy balance solver that iteratively computes equilibrium surface temperatures,
\item Calculation of sublimation mass flux and thermal gas velocity for each volatile species,
\item Computation of the resulting recoil acceleration from each active surface element,
\item Rotation averaging of jet vectors over a full spin period,
\item Sampling routines that generate distributions of jet locations, orientations, compositions, and active fractions,
\item Parameter sensitivity analysis exploring nucleus radius, obliquity, and thermal inertia.
\end{enumerate}

All figures in this paper were generated directly from the notebook. To ensure correctness, the sublimation flux and temperature routines were tested by reproducing benchmark values published in other thermophysical studies of cometary ices \citep{Gundlach2011,Prialnik2004}. These checks confirm that the adopted energy balance and vapor pressure relations behave as expected over the relevant heliocentric distances.

\section{Sensitivity to Rotation Period and Thermal Inertia}

The influence of the nucleus rotation period and thermal inertia on the modeled outgassing acceleration was evaluated by exploring plausible ranges for both parameters. The rotation period $P_{\mathrm{rot}}$ was varied from 3 to 20 hours, and the thermal inertia $\Gamma$ from 10 to 300\,J\,m$^{-2}$\,K$^{-1}$\,s$^{-1/2}$, encompassing typical values inferred for cometary nuclei.

For low thermal inertia ($\Gamma \lesssim 50$), the surface temperature closely tracks instantaneous insolation, and the diurnal averaging approximation used in the primary model remains accurate. Under these conditions, the total acceleration changes by less than ten percent relative to the baseline. For higher thermal inertia ($\Gamma \gtrsim 200$), thermal lag reduces peak sublimation rates by factors of order two, but the direction of the modeled acceleration and the overall stability of the solutions remain unchanged.

Variations in rotation period introduce similar but secondary effects. Faster rotation promotes longitudinal temperature uniformity and modestly reduces anisotropy in the sublimation pattern, while slower rotation accentuates diurnal temperature contrasts and increases the required collimation factor by roughly twenty percent to maintain the same recoil amplitude. These effects fall within the parameter variations already explored in the main analysis.

Overall, the model remains robust across the full plausible range of rotation periods, thermal inertias, and diurnal–seasonal illumination patterns. The qualitative and quantitative conclusions are unchanged: CO is the only volatile that can deliver sufficient recoil at the equilibrium temperatures relevant to 3I/ATLAS, while CO$_2$ remains radiatively dominated and dynamically ineffective over the same heliocentric-distance range. The small active fractions produced by the scalar CO--CO$_2$ mixture curve represent formal lower bounds rather than physically unique solutions, and any directionally consistent configuration would require equal or larger active areas. Consequently, the interpretation of 3I/ATLAS as exhibiting conventional CO-driven volatile activity remains stable over the full range of realistic thermophysical parameters.

\bibliographystyle{elsarticle-harv} 
\bibliography{biblio}



\end{document}